\documentclass[twocolumn,prd,amsmath,amssymb,floatfix]{revtex4}

\usepackage{bbold}
\usepackage{bm}
\usepackage{color}
\usepackage{dcolumn}
\usepackage{feynmf} 
\usepackage{graphics}
\usepackage{graphicx}
\usepackage{subfigure}
\usepackage{slashed}
\usepackage{placeins}

\allowdisplaybreaks
\usepackage{pdfpages}
\usepackage{eso-pic}
\usepackage{everyshi}
\usepackage{pdflscape}

\usepackage{epsfig}

\def\al{\alpha}

\def\veps{\varepsilon}

\newcommand{\ep}{\varepsilon}

\def\be{\begin{equation}}
\def\ee{\end{equation}}
\def\bea{\begin{eqnarray}}
\def\eea{\end{eqnarray}}
\def\bse{\begin{subequations}}
\def\ese{\end{subequations}}
\def\bc{\begin{center}}
\def\ec{\end{center}}

\def\ms{\medskip}

\def\ra{\rightarrow}

\def\nonum{\nonumber}

\def\Ord{{\rm O}}

\begin{document}

\begin{fmffile}{fmfdymNxi}
\fmfcmd{%
vardef cross_bar (expr p, len, ang) =
((-len/2,0)--(len/2,0))
rotated (ang + angle direction length(p)/2 of p)
shifted point length(p)/2 of p
enddef;
style_def crossed expr p =
cdraw p;
ccutdraw cross_bar (p, 5mm,  45);
ccutdraw cross_bar (p, 5mm, -45)
enddef;}

\title{Critical behaviour of ($2+1$)-dimensional QED: \\
       $1/N_f$-corrections in an arbitrary non-local gauge}

\author{A.~V.~Kotikov$^1$ and S.~Teber$^{2,3}$}
\affiliation{
$^1$Bogoliubov Laboratory of Theoretical Physics, Joint Institute for Nuclear Research, 141980 Dubna, Russia.\\
$^2$Sorbonne Universit\'es, UPMC Universit\'e Paris 06, UMR 7589, LPTHE, F-75005, Paris, France.\\
$^3$CNRS, UMR 7589, LPTHE, F-75005, Paris, France.}

\date{\today}

\begin{abstract}
Dynamical chiral symmetry breaking (D$\chi$SB) is studied within ($2+1$)-dimensional QED
with $N$ four-component fermions. The leading and next-to-leading orders of the $1/N$ expansion
are computed exactly. The analysis is carried out in an arbitrary non-local gauge. Resumming the wave-function renormalization constant
at the level of the gap equation yields a strong suppression of the gauge dependence of the critical fermion flavour number, $N_c(\xi)$ where $\xi$ is the gauge fixing parameter, 
which is such that D$\chi$SB takes place for $N<N_c(\xi)$. Neglecting the weak gauge-dependent terms yields $N_c = 2.8469$ while,
in the general case, it is found that: $N_c(1) = 3.0084$ in the Feynman gauge, $N_c(0) = 3.0844$ in the Landau gauge and $N_c(2/3)= 3.0377$ in the $\xi=2/3$ gauge 
where the leading order fermion wave function is finite. These results suggest that D$\chi$SB should take place for integer values $N \leq 3$.
\end{abstract}

\maketitle


\section{Introduction } 

We consider Quantum Electrodynamics in $2+1$ dimensions (QED$_3$) which is described by the Lagrangian: 
\be
L = \overline \Psi ( i \hat \partial - e \hat A ) \Psi - \frac{1}{4} F_{ \mu \nu}^2\, ,
\label{L-QED3}
\ee
where $ \Psi$ is taken to be a four component complex spinor. 
In the presence of $N$ fermion flavours, the model has a $U(2N)$ symmetry. A (parity-invariant) fermion mass term, $m\overline \Psi \Psi$, breaks this symmetry to $U(N) \times U(N)$ (the case of a parity
non-invariant mass will not be considered here). In the massless case, loop expansions are plagued by infrared divergences.
The latter soften upon analysing the model in a $1/N$ expansion \cite{AppelquistP81,JackiwT81+AppelquistH81}.
Since the theory is super-renormalizable, the mass scale is then given by the dimensionful coupling constant: $a = Ne^2/8$, which is kept fixed as $N \rightarrow \infty$. 
Early studies of this model \cite{Pisarski84,AppelquistNW88} suggested that the physics is rapidly damped at momentum scales $p \gg a$ 
and that a (parity-invariant) fermion mass term breaking the flavour symmetry is dynamically generated at scales which are orders of magnitude smaller than 
the intrinsic scale $a$. Since then, dynamical chiral symmetry breaking (D$\chi$SB) in QED$_3$ and the dependence of the dynamical fermion mass on $N$ 
have been the subject of extensive studies, see, {\it e.g.}, 
[\onlinecite{Pisarski84,AppelquistNW88,Pennington91+92,Pisarski91,AtkinsonJM90,KarthikN16,DagottoKK89+90,Azcoiti93+96,AppelquistCS99,GiombiKT16,DiPietroKSS16,Braun:2014wja,Janssen:2016nrm,Nash89,Kotikov93+12,KotikovST16,Herbut16,Giombi:2016fct,Gusynin:2016som,BashirRSR09}].

A central issue is related to the value of the critical fermion number, $N_c$, which is such that D$\chi$SB takes place only for $N<N_c$. 
An accurate determination of $N_c$ is of crucial importance to understand the phase structure of QED$_3$ with far reaching implications from particle physics 
to planar condensed matter physics systems having relativistic-like low-energy excitations such as some two-dimensional antiferromagnets \cite{MarstonA89+IoffeL89}
and graphene \cite{Semenoff84+Wallace47}. It turns out that the values that can be found in the literature vary from $N_c \ra \infty$ \cite{Pisarski84,Pennington91+92,Pisarski91,Azcoiti93+96} 
corresponding to D$\chi$SB for all values of $N$, all the way to $N_c \ra 0$ in the case where no sign of D$\chi$SB is found \cite{AtkinsonJM90,KarthikN16}.
Recent works based on conformal field theory techniques tend to narrow this range but the upper bound found for $N_c$ still varies: $N_c <3/2$ \cite{AppelquistCS99} or
$N_c < 4.4$ \cite{GiombiKT16} or $N_c<9/4$ \cite{DiPietroKSS16}. Other works suggest that there might be two different critical flavor
numbers \cite{Braun:2014wja,Janssen:2016nrm}: $N_c < N_c^{\text{conf}}$ where $N_c \approx 4$ \cite{Braun:2014wja}, $N_c \leq 4.422$ \cite{Janssen:2016nrm} 
and above $4.1<N_c^{\text{conf}}<10.0$ \cite{Braun:2014wja}, $N_c^{\text{conf}} \approx 6.24$ \cite{Janssen:2016nrm}, the theory is quasi-conformal.
Of importance to us in the following, is the approach of Appelquist et al.\ \cite{AppelquistNW88} who found that $N_c = 32/ \pi^2 \approx 3.24$ by solving 
the Schwinger-Dyson (SD) gap equation using a leading order (LO) $1/N$-expansion. Lattice simulations in agreement with a finite non-zero value of $N_c$ can be found in [\onlinecite{DagottoKK89+90}]. 
Soon after the analysis of [\onlinecite{AppelquistNW88}], Nash approximately included next-to-leading order (NLO) corrections and performed a partial resummation
of the wave-function renormalization constant at the level of the gap equation; he found \cite{Nash89}: $N_c \approx 3.28$. Recently, upon refining the work of [\onlinecite{Kotikov93+12}], 
the NLO corrections could be computed exactly in the Landau gauge yielding (in the absence of resummation) \cite{KotikovST16}: $N_c \approx 3.29$, a value which is surprisingly close to the one of Nash 
\footnote{We have been informed by V.~Gusynin, see also \cite{Gusynin:2016som}, that
Eq.~(15) in Ref.~[\onlinecite{Nash89}] contains an error: ``341'' should be replaced by ``277'' which then leads to $N_c=3.52$.}. 
More recently, using different methods, several new estimates were given: $N_c =1 + \sqrt{2} \approx 2.41$ in \cite{Giombi:2016fct}, $N_c \approx 2.89$ in \cite{Herbut16} and 
$N_c \approx 2.85$ in \cite{Gusynin:2016som}. 

The purpose of the present work is to extend the exact results of [\onlinecite{KotikovST16}] to an arbitrary non-local gauge [\onlinecite{Simmons90+KugoM92}].
Such an achievement represents an essential improvement with respect to Nash's approximate NLO results which were carried out in the Feynman gauge 27 years ago.
In this respect, there is presently a strong ongoing interest in the study of the gauge dependence of D$\chi$SB in
several models, see [\onlinecite{AhmadCCR16,BashirRSR09}]. The choice of the Landau gauge in [\onlinecite{KotikovST16}] 
was motivated by recent results on QED$_3$ \cite{BashirRSR09} 
showing the gauge invariance of $N_c$ in this gauge when using the Ball-Chiu vertex \cite{BallC80}.
Actually, after resumming the wave-function renormalization constant, we find that the LO term in the gap equation becomes gauge-invariant, in agreement with Nash, 
but also that NLO terms become only weakly gauge-variant. As will be shown in the following, this leads to a  very stable value of $N_c$ upon varying the gauge-fixing parameter.
Moreover, the large-$N$ limit of the photon propagator in QED$_3$ has precisely the same momentum dependence as the one 
in the so-called reduced QED, see [\onlinecite{GorbarGM01}] and also [\onlinecite{Marino93+DoreyM92+KovnerR90}]. 
One difference is that the gauge fixing parameter in reduced QED is twice less than the one in QED$_3$. 
Such a difference can be taken into account with the help of our present results for QED$_3$ together with the multi-loop results obtained in [\onlinecite{Teber12+KotikovT13,KotikovT14}]. 
The case of reduced QED, and its relation with dynamical gap generation in graphene which is the subject of active ongoing research, see, {\it e.g.},
the reviews [\onlinecite{KotovUPGC12,MiranskyS15}], will be considered in a separate paper \cite{Kotikov:2016yrn}. In the following, we shall focus exclusively on QED$_3$.

The paper is organized as follows. In Sec.~\ref{sec:LO}, the LO results are presented and in Sec.~\ref{sec:NLO}
the NLO ones including Nash's resummation are presented. In Sec.~\ref{sec:conclusion} the conclusion is given and
in App.~\ref{app} some details related to the resummation procedure within our working frame are provided.

\section{Schwinger-Dyson equations and Leading Order} 
\label{sec:LO}

With the conventions of Ref.~[\onlinecite{KotikovST16}], the inverse fermion propagator is defined as:
$S^{-1}(p) = [1 + A(p)]\,\left( i\hat p + \Sigma (p) \right)$ 
where $A(p)$ is the fermion wave function  and $\Sigma (p)$ is the dynamically generated parity-conserving mass which 
is taken to be the same for all the fermions. The SD equation for the fermion propagator may be decomposed into scalar and vector components
as follows:
\begin{subequations}
\label{SD-sigma+A}
\begin{eqnarray}
\tilde{\Sigma}(p)
= \frac{2a}{N} \, \mbox{Tr} \int \frac{d^3 k}{(2 \pi )^3}
\frac{\gamma^{\mu} D_{\mu \nu}(p-k) \Sigma (k) \Gamma^{\nu}(p,k)}
{\left[1 + A(k) \right] \left( k^2 + \Sigma^2(k) \right)}\, ,  
\label{SD-sigma} \\
A(p) p^2 =  -\frac{2a}{N} \, \mbox{Tr} \int \frac{d^3 k}{(2 \pi )^3}
\frac{ D_{\mu \nu}(p-k) \hat p \gamma^{\mu} \hat k \Gamma^{\nu}(p,k)}
{\left[1 + A(k) \right] \left( k^2 + \Sigma^2(k) \right)} \, ,
\label{SD-A}
\end{eqnarray}
\end{subequations}
where  $\tilde{\Sigma} (p) = \Sigma (p)[1 + A(p)]$, 
$D_{\mu \nu}(p)$ is the photon propagator in the non-local $\xi$-gauge:
\be
D_{\mu \nu}(p) = \frac{P_{\mu \nu}^\xi(p)}{p^2 \left[1 + \Pi (p) \right]}, \quad P_{\mu \nu}^\xi(p) = g_{\mu \nu} - (1-\xi)\frac{p_\mu p_\nu}{p^2}\, ,
\label{photon}
\ee
$\Pi(p)$ is the polarization operator and $\Gamma^{\nu}(p,k)$ is the vertex function.
In the following, Eqs.~(\ref{SD-sigma+A}) will be studied for an arbitrary value of the gauge-fixing parameter $\xi$.
All calculations will be performed with the help of the standard rules of perturbation theory for massless Feynman diagrams
as in [\onlinecite{Kazakov83}], see also the recent short review [\onlinecite{TeberK16}]. 
For the most complicated diagrams, the Gegenbauer polynomial technique will be used following [\onlinecite{Kotikov95}].

\begin{figure}[tl]
    \includegraphics[width=0.18\textwidth]{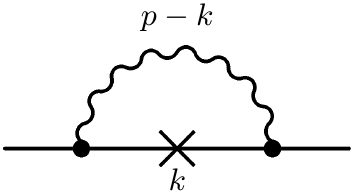}
    \caption{\label{fig:diags-LO}
        LO diagram to the dynamically generated mass $\Sigma(p)$. The crossed line denotes mass insertion.}
\end{figure}
%


To start with, let us consider the LO approximations in the $1/N$ expansion. The latter are given by:
$A(p) = 0$, $\Pi (p) = a/ |p|$ and $\Gamma^{\nu}(p,k) = \gamma^{\nu}$,
%
%
where the fermion mass has been neglected~\footnote{A study of the fermion mass contribution to $\Pi(p)$ can be
found, for example, in [\onlinecite{GusyninHR96}].} in the calculation of $\Pi(p)$. 
A single diagram contributes to the gap equation (\ref{SD-sigma}) at LO, see Fig.~\ref{fig:diags-LO}, and the latter reads:
\be
  \Sigma (p) =\frac{8(2+\xi)a}{N}
   \int \frac{d^3 k}{(2 \pi )^3}
\frac{ \Sigma (k) }{ \left( k^2 + \Sigma^2(k)
\right)
\bigl[ (p-k)^2 + a \,|p-k| \bigr]} \, . 
\label{SD-sigma-LO1} 
\ee
Following [\onlinecite{AppelquistNW88}], we consider the limit of large $a$ and linearize Eq.~(\ref{SD-sigma-LO1})
which yields:
\be
  \Sigma (p) =
 \frac{8(2 + \xi)}{N}
   \int \frac{d^3 k}{(2 \pi )^3} \frac{ \Sigma (k) }{k^2 \, |p-k| } \, .
\label{SD-sigma-LO4}
\ee
The mass function may then be parametrized as \cite{AppelquistNW88}:
\be
\Sigma (k) = B \, (k^2)^{ -\alpha} \, ,
\label{sigma-parametrization}
\ee
where $B$ is arbitrary and the index $\al$ has to be self-consistently determined.
Using this ansatz, Eq.~(\ref{SD-sigma-LO4}) reads:
\be
\Sigma^{(\text{LO})}(p) = \frac{4(2+\xi)B}{N}\,\frac{(p^2)^{-\al}}{(4\pi)^{3/2}}\, \frac{2\beta}{\pi^{1/2}}\, ,
\label{sigma-LO-res}
\ee
from which the LO gap equation is obtained:
\bea
1 = \frac{(2+\xi)\beta}{L},~~ \text{where} ~~ \beta = \frac{1}{\alpha \left( 1/2 - \alpha \right)} ~~ \text{and} ~~  L \equiv \pi^2 N\, . 
\label{gap-eqn-LO}
\eea
%
%
%
Solving the gap equation, yields:
\begin{eqnarray}
\alpha_{\pm} = \frac{1}{4}\,\left( 1 \pm \sqrt{1 - \frac{16(2+\xi)}{L}} \right) \, ,
\label{al-LO}
\end{eqnarray}
which reproduces the solution given by Appelquist et al.\ [\onlinecite{AppelquistNW88}].
The gauge-dependent critical number of fermions: $N_c \equiv N_c(\xi) = 16(2+\xi)/ \pi^2$,
is such that $\Sigma(p) = 0$ for $N>N_c$ and:
\be
\Sigma(0) \simeq \exp \bigl[ - 2 \pi / (N_c/N - 1)^{1/2} \bigr]\, , 
\ee
for $N<N_c$. Thus, D$\chi$SB occurs when $\alpha$ becomes complex, that is for $N<N_c$.

The gauge-dependent fermion wave function may be computed in a similar way. At LO, Eq.~(\ref{SD-A}) simplifies as:
\be
  A(p) p^2 = -\frac{2a}{N}
 \text{Tr} \int \! \! \frac{d^D k}{(2 \pi )^D}
\frac{ P_{\mu \nu}^\xi(p-k) \hat p \gamma^{\mu} \hat k \gamma^{\nu}}
{k^2 |p-k|}\, , 
\label{SD-A1}
\ee
where the integral has been dimensionally regularized with $D=3 -2 \veps$. 
Taking the trace and computing the integral on the r.h.s.\ yields:
\be
  A(p) =
\frac{\Gamma(1+\ep)(4\pi)^{\ep} \mu^{2\ep}}{p^{2\ep}} \, C_1(\xi) = 
  \frac{\overline{\mu}^{2\ep}}{p^{2\ep}} \, C_1(\xi) \, + \Ord(\ep) \, ,
\label{SD-A2}
\ee
where the $\overline{MS}$ parameter $\overline{\mu}$ has the standard form
$\overline{\mu}^2 = 4\pi e^{-\gamma_E} \mu^2$ with the Euler constant $\gamma_E$ and
\be
C_1(\xi)=
+\frac{2}{3\pi^2N} \left((2-3\xi)\left[\frac{1}{\ep} - 2\ln 2\right]+ \frac{14}{3} - 6\xi\right)\, .
\label{C1}
\ee
We note that in the $\xi=2/3$-gauge, the value of $A(p)$ is finite and
$C_1(\xi=2/3)= +4/(9\pi^2N)$.
From Eqs.~(\ref{SD-A2}) and (\ref{C1}), the LO wave-function renormalization constant may be extracted:
\be
\lambda_A = \mu \frac{d A(p)}{d\mu} = \frac{4(2 - 3\xi)}{3\pi^{2}N}\, , 
\label{lambdaA-LO}
\ee
a result which coincides with the one of [\onlinecite{Gracey94}].


%

\section{Next-to-leading order} 
\label{sec:NLO}

\subsection{Self-energy contributions}

\begin{widetext}

\begin{figure}[t]
    \includegraphics[width=0.96\textwidth]{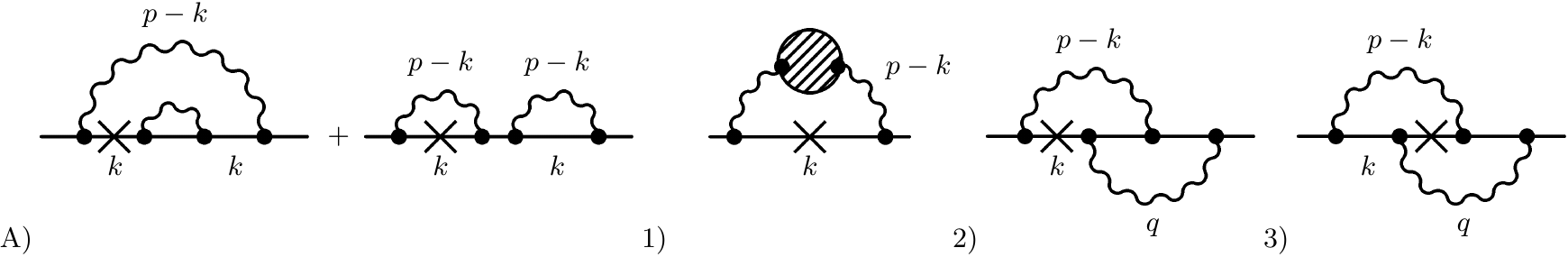}
    \caption{\label{fig:diags-NLO}
        NLO diagrams to the dynamically generated mass $\Sigma(p)$. The shaded blob is defined in Eq.~(\ref{polar-2loops}).}
\end{figure}

\end{widetext}

We now consider the NLO contributions and parametrize them as:
\be
\Sigma^{(\text{NLO})}(p) = \left(\frac{8}{N}\right)^2 B\,\frac{(p^2)^{-\al}}{(4\pi)^{3}}\,
\left( \Sigma_A + \Sigma_1 + 2\,\Sigma_2 + \Sigma_3 \right) \, ,
\ee
where each contribution to the linearized gap equation is represented graphically in Fig.~\ref{fig:diags-NLO}.
Adding these contributions to the LO result, Eq.~(\ref{sigma-LO-res}), 
the gap equation has the following general form:
\begin{eqnarray}
1 = \frac{(2+\xi) \beta}{L} +
\frac{\overline{\Sigma}_A (\xi) + \overline{\Sigma}_1(\xi) + 2\,\overline{\Sigma}_2(\xi) + \overline{\Sigma}_3(\xi)}{L^2} \, ,
\label{gap-eqn-NLO}
\end{eqnarray}
where $\overline{\Sigma}_i = \pi \Sigma_i$, $(i=1,2,3.A)$.
In [\onlinecite{KotikovST16}], these contributions were computed in the Landau gauge, $\xi=0$. After very tedious and lengthy calculations, 
these computations could be extended to an arbitrary non-local gauge. We now summarize our results.

The contribution  $\Sigma_A$, see Fig.~\ref{fig:diags-NLO} A), originates from the LO value of $A(p)$ and is singular.
Using dimensional regularization, and for an arbitrary parameter $\xi$, it reads:
\bea
&&\overline{\Sigma}_A(\xi) = 4 \,
\frac{\overline{\mu}^{2\ep}}{p^{2\ep}}\, \beta 
\left[ \left(\frac{4}{3}(1-\xi)-\xi^2\right)\left[
\frac{1}{\ep} + \Psi_1 - \frac{\beta}{4} \right] \right .
\nonum \\
&&\left . + \left(\frac{16}{9}-\frac{4}{9}\xi-2\xi^2\right)
\right] \, ,
\label{sigma-NLO-A.1}
\eea
where 
\be
\Psi_1 = \Psi(\alpha)+  \Psi(1/2-\alpha)-2\Psi(1) + \frac{3}{1/2-\alpha} 
-2 \ln 2\, ,
\label{psi1}
\ee
and $\Psi$ is the digamma function. 

The contribution of diagram 1) in Fig.~\ref{fig:diags-NLO} is finite and reads:
\be
\overline{\Sigma}_1(\xi) = -2(2+\xi)\, \beta\, \hat{\Pi}, \qquad \hat{\Pi} = \frac{92}{9} - \pi^2\, ,
\label{sigma-NLO-1}
\ee
where the gauge dependence comes from the fact that we work in a non-local gauge
and $\hat{\Pi}$ arises from the two-loop polarization operator in dimension $D=3$~\cite{Gracey93,GusyninHR01,Teber12+KotikovT13} which may be graphically represented as:
\be
\parbox{8mm}{
    \begin{fmfgraph*}(8,7)
      \fmfleft{i}
      \fmfright{o}
      \fmfleft{ve}
      \fmfright{vo}
      \fmffreeze
      \fmfforce{(-0.3w,0.5h)}{i}
      \fmfforce{(1.3w,0.5h)}{o}
      \fmfforce{(0w,0.5h)}{ve}
      \fmfforce{(1.0w,0.5h)}{vo}
      \fmffreeze
      \fmf{photon}{i,ve}
      \fmf{photon}{vo,o}
      \fmffreeze
      \fmfdot{ve,vo}
      \fmf{phantom,tag=1}{ve,vo}
      \fmfposition
      \fmfipath{p[]}
      \fmfiset{p1}{vpath1(__ve,__vo)}
      \def\blob#1{%
        \fmfiv{decor.shape=circle,decor.filled=shaded,decor.size=1w}{#1}}
      \blob{point length(p1)/2 of p1}
    \end{fmfgraph*}
} \qquad = \quad 2 \times ~
\parbox{8mm}{
    \begin{fmfgraph*}(14,8)
      \fmfleft{i}
      \fmfright{o}
      \fmf{photon}{i,v1}
      \fmf{photon}{v2,o}
      \fmf{phantom,right,tension=0.1,tag=1}{v1,v2}
      \fmf{phantom,right,tension=0.1,tag=2}{v2,v1}
      \fmf{phantom,tension=0.1,tag=3}{v1,v2}
      \fmfdot{v1,v2}
      \fmfposition
      \fmfipath{p[]}
      \fmfiset{p1}{vpath1(__v1,__v2)}
      \fmfiset{p2}{vpath2(__v2,__v1)}
      \fmfiset{p3}{vpath3(__v1,__v2)}
      \fmfi{plain}{subpath (0,length(p1)) of p1}
      \fmfi{plain}{subpath (0,length(p2)/4) of p2}
      \fmfi{plain}{subpath (length(p2)/4,3length(p2)/4) of p2}
      \fmfi{plain}{subpath (3length(p2)/4,length(p2)) of p2}
      \fmfi{photon}{point length(p2)/4 of p2 .. point length(p3)/2 of p3 .. point 3length(p2)/4 of p2}
      \def\vert#1{%
        \fmfiv{decor.shape=circle,decor.filled=full,decor.size=2thick}{#1}}
      \vert{point length(p1)/2 of p1}
      \vert{point length(p2)/2 of p2}
    \end{fmfgraph*}
} \qquad + \quad
\parbox{8mm}{
    \begin{fmfgraph*}(14,8)
      \fmfleft{i}
      \fmfright{o}
      \fmf{photon}{i,v1}
      \fmf{photon}{v2,o}
      \fmf{phantom,right,tension=0.1,tag=1}{v1,v2}
      \fmf{phantom,right,tension=0.1,tag=2}{v2,v1}
      \fmf{phantom,tension=0.1,tag=3}{v1,v2}
      \fmfdot{v1,v2}
      \fmfposition
      \fmfipath{p[]}
      \fmfiset{p1}{vpath1(__v1,__v2)}
      \fmfiset{p2}{vpath2(__v2,__v1)}
      \fmfi{plain}{subpath (0,length(p1)/2) of p1}
      \fmfi{plain}{subpath (length(p1)/2,length(p1)) of p1}
      \fmfi{plain}{subpath (0,length(p2)/2) of p2}
      \fmfi{plain}{subpath (length(p2)/2,length(p2)) of p2}
      \fmfi{photon}{point length(p1)/2 of p1 -- point length(p2)/2 of p2}
      \def\vert#1{%
        \fmfiv{decor.shape=circle,decor.filled=full,decor.size=2thick}{#1}}
      \vert{point length(p1)/2 of p1}
      \vert{point length(p2)/2 of p2}
    \end{fmfgraph*}
} \qquad .
\label{polar-2loops}
\ee
\ms

The contribution of diagram 2) in Fig.~\ref{fig:diags-NLO} is again singular. Dimensionally regularizing it yields:
\begin{subequations}
\bea
&&\overline{\Sigma}_2 (\xi)= -2\,\frac{\overline{\mu}^{2\ep}}{p^{2\ep}}\, \beta \, 
\bigg[ \frac{(2+\xi)(2-3\xi)}{3}\,\left(\frac{1}{\veps} + \Psi_1 - \frac{\beta}{4} \right) \bigg .
\label{sigma-NLO-22} \\
&&\bigg . + \frac{\beta}{4}\,\left(\frac{14}{3}\,(1-\xi) + \xi^2 \right) 
+ \frac{28}{9} + \frac{8}{9}\,\xi - 4 \xi^2 \bigg] + (1-\xi)\,\hat{\Sigma}_{2} \, ,
\nonum\\
&&\hat{\Sigma}_2(\alpha) = (4\alpha-1) \beta \Bigl[\Psi'(\alpha) - \Psi'(1/2-\alpha)\Bigl] 
\nonum \\
&&+\frac{\pi}{2\alpha}\,\tilde{I}_1(\alpha) + \frac{\pi}{2(1/2-\alpha)} \tilde{I}_1(\alpha+1) \, ,
\label{sigma-NLO-22_L} 
\eea
\end{subequations}
where $\Psi'$ is the trigamma function and $\tilde{I}_1(\alpha)$ is a dimensionless integral that was defined in [\onlinecite{KotikovST16}].

The singularities in $\overline{\Sigma}_A(\xi)$ and $\overline{\Sigma}_2(\xi)$ cancel each other and their sum is therefore finite.
Defining: $\overline{\Sigma}_{2A} (\xi)= \overline{\Sigma}_{A} (\xi) + 2\overline{\Sigma}_{2} (\xi)$, the latter reads:
\bea
&&\overline{\Sigma}_{2A}(\xi) =  2(1-\xi) \hat{\Sigma}_2(\alpha) - \left(\frac{14}{3}(1-\xi)+\xi^2\right) \beta^2 
\nonum \\
&& - 8\beta\left(\frac{2}{3}(1+\xi)-\xi^2\right) \, .
\eea

Finally, the contribution of diagram 3) in Fig.~\ref{fig:diags-NLO} is finite and reads:
\begin{subequations}
\bea
&&\overline{\Sigma}_3(\xi) = \hat{\Sigma}_3(\alpha,\xi) + \Bigl(3+4\xi-2\xi^2) \beta^2\, , 
 \\
&&\hat{\Sigma}_3(\alpha,\xi) = \frac{1}{4}\bigl(1+8\xi+\xi^2+2\alpha (1-\xi^2)\bigr)
\pi\tilde{I}_2(\alpha)
\nonum \\
&&+ \frac{1}{2}\bigl(1+4\xi-\alpha (1-\xi^2)\bigr)
\pi\tilde{I}_2(1+\alpha) 
\nonum \\
&&+ \frac{1}{4}\bigl(-7-16\xi+3\xi^2\bigr)
\pi\tilde{I}_3(\alpha)\, ,
\label{sigma-NLO-3AG}
\eea
\end{subequations}
where the dimensionless integrals $\tilde{I}_2(\alpha)$ and $\tilde{I}_3(\alpha)$
were defined in [\onlinecite{KotikovST16}].

\subsection{Gap equation (1)} 

Combining all of the above results, the gap equation (\ref{gap-eqn-NLO}) may be written in an explicit form as:
\bea
&&1 = \frac{(2+\xi)\beta}{L} + \frac{1}{L^2}\,
\Bigl[8 S(\al,\xi) - 2(2+\xi) \hat{\Pi} \beta \Bigl . 
\nonum \\
&&\Bigr . + \left(-\frac{5}{3}+ \frac{26}{3}\xi -3\xi^2\right) \beta^2 - 8\beta\left(\frac{2}{3}(1-\xi)-\xi^2\right) \Bigr]\, , \qquad
\label{gap-eqn-NLO-explicit}
\eea
where 
\be
S(\al,\xi) = \Bigl(\hat{\Sigma}_3(\al,\xi)+2(1-\xi) \hat{\Sigma}_2(\alpha)\Bigr)/8 \, .
\label{delta}
\ee
%


At this point, we consider Eq.~(\ref{gap-eqn-NLO-explicit}) directly at the critical point $\alpha=1/4$, {\it i.e.}, at $\beta=16$. 
This yields:
\bea
&&L_c^2 -16(2+\xi) L_c - 8\left [ S(\xi) - 4(2+\xi) \hat{\Pi} \right .
\nonum \\
&& \left . - 16 \left(4- 50 \xi / 3 + 5\xi^2\right) \right ] = 0 \, ,
 \label{Lc-eqn}
\eea
where $S(\xi)= S(\al=1/4,\xi)= (\hat{\Sigma}_3(\xi)+2(1-\xi) \hat{\Sigma}_2)/8$ with $\hat{\Sigma}_2 = \hat{\Sigma}_2(\al=1/4)$ and 
$\hat{\Sigma}_3(\xi)= \hat{\Sigma}_3(\al=1/4,\xi)$. Solving Eq.~(\ref{Lc-eqn}), we have two standard solutions:
%
\bea
&&L_{c,\pm} = 8(2+\xi) \pm \sqrt{d_1(\xi)}\, ,
\label{Lc-solutions} \\ 
&&d_1(\xi)= 8\left [ S(\xi)-8\left(4-\frac{112}{3}\xi+9\xi^2 + \frac{2+\xi}{2} \hat{\Pi} \right) \right ] \, .
\nonum 
\eea
%
In order to provide a numerical estimate for $N_c$, we have used the series representations~\cite{KotikovST16} to evaluate the integrals: 
$\pi \tilde{I}_1(\alpha=1/4) \equiv R_1$ and $\pi \tilde{I}_2(\alpha=1/4 + i\delta) \equiv R_2 - iP_2  \delta + O(\delta^2)$ where $\delta \ra 0$ regulates an artificial singularity in $\pi \tilde{I}_3(\alpha=1/4)=R_2 + P_2/4$.
With 10000 iterations for each series, the following numerical estimates are obtained \cite{KotikovST16}: 
\be
R_1=163.7428, \quad R_2=209.175, \quad P_2=1260.720 \, ,
\label{Is-numerics}
\ee
from which the complicated part of the self-energies can be evaluated:
\be
\hat{\Sigma}_2= 4 R_1, \quad \hat{\Sigma}_3(\xi)=(\xi^2-1) R_2 - (7+16 \xi -3 \xi^2)\,P_2 / 16\, .
\label{sigma2.3}
\ee
Combining these values with the one of $\hat{\Pi}$, yields: 
\be
N_c(\xi=0)=3.29, \quad N_c(\xi=2/3)=3.09\, ,
\label{Nc-no-resum}
\ee
where ``$-$'' solutions are unphysical and there is no solution in the Feynman gauge.
In the Landau gauge, we recover the result of \cite{KotikovST16}.
%
%
The range of $\xi$-values for which there is a solution corresponds to $\xi_- \leq  \xi \leq  \xi_+$, where $\xi_{+}=0.88$ and  $\xi_{-}=-2.36$. 

\subsection{Gap equation (2)} 

Following Ref.~[\onlinecite{Nash89}], we would like to resum the LO term together with part of the NLO corrections containing terms $\sim \beta^2$.
In order to do so, we will now rewrite the gap equation (\ref{gap-eqn-NLO-explicit}) in a form which is suitable for resummation.
This amounts to extract the terms  $\sim \beta$ and $\sim \beta^2$ from the complicated parts of the fermion self-energy, Eqs.~(\ref{sigma-NLO-22_L}) and (\ref{sigma-NLO-3AG}).
All calculations done this yields:
\begin{subequations}
\label{tSigma2}
\bea
&&\hat{\Sigma}_2(\alpha) = \beta \bigl(3\beta -8 \bigr) + \tilde{\Sigma}_2(\alpha)\, , 
\label{Sigma2} \\
&&\tilde{\Sigma}_2 = \tilde{\Sigma}_2(\alpha=1/4)=4 \tilde{R}_1,~~ \tilde{R}_1=3.7428\, ,
\label{tSigma2-1/4} 
\eea
\end{subequations}
where the rest, $\tilde{\Sigma}_2 = \tilde{\Sigma}_2(\alpha=1/4)$, was determined by imposing Eq.~(\ref{sigma2.3}). Similarly:
%
\begin{subequations}
\label{tSigma3}
\bea
&&\overline{\Sigma}_3(\xi) = -4 \xi (4+\xi) \beta + \tilde{\Sigma}_3(\alpha,\xi)\, ,  
\label{tsigma-NLO-3AG0} \\
&&\tilde{\Sigma}_3(\alpha,\xi) = \frac{1}{4}\bigl(1+8\xi+\xi^2+2\alpha (1-\xi^2)\bigr)
\pi\tilde{J}_2(\alpha)
\nonum \\
&&+ \frac{1}{2}\bigl(1+4\xi-\alpha (1-\xi^2)\bigr)
\pi\tilde{J}_2(1+\alpha) 
\nonum \\
&&- \frac{1}{4}\bigl(-7-16\xi+3\xi^2\bigr)
\pi\tilde{J}_3(\alpha)\, ,
\label{tsigma-NLO-3AG} \\
&&\tilde{\Sigma}_3(\xi) = \bigl(\xi^2-1\bigr)  \tilde{R}_2 -  \bigl(7+16\xi -3\xi^2 \bigr)\,\frac{\tilde{P}_2}{16}\, ,
\label{tR2}
\eea
\end{subequations}
where the form of the rest, $\tilde{\Sigma}_3(\xi)=\tilde{\Sigma}_3(\alpha=1/4,\xi)$, is imposed by Eq.~(\ref{sigma2.3}). 
Equating Eqs.~(\ref{tR2}) with (\ref{tsigma-NLO-3AG}) for $\al=1/4$ together with using the values:
\begin{subequations}
\label{tJ}
\begin{flalign}
&\pi \tilde{J}_2(\alpha=1/4) = \pi \tilde{J}_2(\alpha=5/4) = \tilde{R}_2^N = 17.175\, , 
\\
&\pi \tilde{J}_3(\alpha=1/4) = \tilde{R}_2^N + \tilde{P}_2^N/4\, ,
\\
&\tilde{P}_2^N = -19.28\, , 
\end{flalign}
\end{subequations}
yields:
%
%
%
\be
\tilde{R}_2 = \tilde{R}_2^N -16 = 1.175,~~ \tilde{P}_2 = \tilde{P}_2^N = -19.28\, .
\label{til}
\ee
With the help of the results (\ref{tSigma2}) and (\ref{tSigma3}), the gap equation (\ref{gap-eqn-NLO-explicit}) may be written as:
\bea
&&1 = \frac{(2+\xi)\beta}{L} + \frac{1}{L^2}
\Bigl[8 \tilde{S}(\al,\xi) -2(2+\xi) \hat{\Pi} \beta  \Bigr .
\nonum \\
&&\Bigl . + \left(\frac{2}{3}-\xi\right)\bigl(2+\xi\bigr)\, \beta^2 +
4\beta\left(\xi^2-\frac{4}{3}\xi - \frac{16}{3}\right) \Bigr]\, , \quad
 \label{gap-eqn-NLO-explicit_1} 
\eea
%
where 
\be
\tilde{S}(\al,\xi) = \Bigl(\tilde{\Sigma}_3(\al,\xi)+2(1-\xi) \tilde{\Sigma}_2(\al) \Bigr)/8 \, .
\label{tdelta} 
\ee
At this point Eqs.~(\ref{gap-eqn-NLO-explicit}) and (\ref{gap-eqn-NLO-explicit_1}) are strictly equivalent to each other and yield the same values for $N_c(\xi)$.

\subsection{Resummation} 

Eq.~(\ref{gap-eqn-NLO-explicit_1}) is the convenient starting point to perform a resummation of the wave function renormalization constant.
Up to second order, the expansion of the latter reads:
\be
\lambda_A = \frac{\lambda^{(1)}}{L} + \frac{\lambda^{(2)}}{L^2} + \cdots, \quad \lambda^{(1)} = 4 \left(\frac{2}{3}-\xi \right)\, ,
\label{lambdaA}
\ee
where $\lambda^{(1)}$ is the LO part and $\lambda^{(2)}$ the NLO one. The latter can be obtained
from Gracey's calculations~\cite{Gracey93} and reads:
\bea
\lambda^{(2)} = - 8 \left(\frac{8}{27} + \left(\frac{2}{3}-\xi \right) \hat{\Pi} \right)\, .
\label{lambdaA-NLO}
\eea
As can be seen from Eq.~(\ref{gap-eqn-NLO-explicit_1}), the NLO term $\sim \beta^2$ is proportional to the LO wave function renormalization constant. This term, together
with the LO term in the gap equation, can be thought of the first and zeroth order terms, respectively, of an expansion in $\lambda_A$.
Following Nash, it is possible to resum the full expansion of $\lambda_A$ at the level of the gap equation (see Appendix for details) leading to:
\be
1 = \frac{8\beta}{3L} +   \frac{\beta}{4L^2} \left(\lambda^{(2)} - 4\lambda^{(1)}   
\left(\frac{14}{3}+\xi \right) \right) + \frac{\Delta(\al,\xi)}{L^2} \, ,
\label{gap-eqn-NLO-explicit_5a} 
\ee
where 
\be
\Delta(\al,\xi) = 8 \tilde{S}(\al,\xi) - 4\beta\,(\xi^2 +4\xi + 8/3)  - 2\beta\,(2+\xi)\,\hat{\Pi}\, .
\ee
Interestingly, the LO term in Eq.~(\ref{gap-eqn-NLO-explicit_5a}) is now gauge independent. With the help of 
Eqs.~(\ref{lambdaA}) and (\ref{lambdaA-NLO}), Eq.~(\ref{gap-eqn-NLO-explicit_5a}) can now be explicited as:
%
\bea
1 = \frac{8\beta}{3L}  + \frac{1}{L^2}\,\Bigl[ 8 \tilde{S}(\al,\xi) - \frac{16}{3} \, \beta \, \left(\frac{40}{9} + \hat{\Pi} \right) \Bigr]\, ,
\label{gap-eqn-NLO-explicit_5b}
\eea
%
which displays a strong suppression of the gauge dependence even at NLO as $\xi$-dependent terms do exist but they enter the gap equation only through the rest, $\tilde{S}$, which is very small numerically.

We now consider Eq.~(\ref{gap-eqn-NLO-explicit_5b}) at the critical point, $\al=1/4$ ($\beta=16$), which yields:
\bea
L_c^2 -\frac{128}{3} L_c - \Bigl[ 8 \tilde{S}(\xi) - \frac{256}{3}\,\left(\frac{40}{9} + \hat{\Pi} \right) \Bigr] = 0 \, .
\label{Lc-eqn_1}
\eea
%
Solving Eq.~(\ref{Lc-eqn_1}), we have two standard solutions:
\begin{subequations}
\bea
&&L_{c,\pm} = \frac{64}{3} \pm \sqrt{d_2(\xi)} \, ,
\label{Lc-solutions_1} \\
&&d_2(\xi)= {\left(\frac{64}{3}\right)}^2 + \Bigl[ 8 \tilde{S}(\xi) - \frac{256}{3}\,\left(\frac{40}{9} + \hat{\Pi} \right) \Bigr]\, . \quad
\label{d2}
\eea
\end{subequations}
In order to provide a numerical estimate for $N_c$,  we have used the values of $\tilde{R}_1$, $\tilde{R}_2$ and $\tilde{P}_2$ of Eqs.~(\ref{tSigma2}) and (\ref{til}).
Combining these values with: 
%
\begin{subequations}
\bea
&&\tilde{S}(\xi=0)=\tilde{R}_1-\frac{\tilde{R}_2}{8}-\frac{7\tilde{P}_2}{128}\, , 
\\ 
&&\tilde{S}(\xi=1)= -\frac{5\tilde{P}_2}{32}\, ,
\\
&&\tilde{S}(\xi=2/3)=\frac{\tilde{R}_1}{3}-\frac{5\tilde{R}_2}{72} - \frac{49 \tilde{P}_2}{384}\, ,
\eea
\end{subequations}
together with the value of $\hat{\Pi}$, yields, for $L_c(\xi)$ and $N_c(\xi)$ (``$-$'' solutions being unphysical): 
\begin{subequations}
\bea
&&L_c(0)=30.44, ~ L_c(2/3)=29.98, ~ L_c(1)=29.69\, , \qquad
\label{Lc-res} \\
&&N_c(0)=3.08, ~ N_c(2/3)=3.04, ~ N_c(1)=3.01\, . \qquad
\label{Nc-values}
\eea
\end{subequations}
Actually, solutions exist for a broad range of values of $\xi$: $\tilde{\xi}_- \leq  \xi \leq  \tilde{\xi}_+$,
where $\tilde{\xi}_{+}=4.042$ and  $\tilde{\xi}_{-}=-8.412$; this is consistent with the weak $\xi$-dependence of the gap equation. 
Moreover, following [\onlinecite{GorbarGM01}], we think that the ``right(est)'' gauge choice is one close to $\xi=2/3$ where the LO fermion wave function is finite.
Indeed, as can be seen by comparing Eqs.~(\ref{Nc-no-resum}) and (\ref{Nc-values}), upon resumming the theory, the value  of $N_c(\xi)$ increases (decreases) for small (large)
values of $\xi$. For $\xi=2/3$, the value of $N_c$ is very stable, decreasing
only by $1$-$2\%$ during resummation. Finally, if we neglect the rest, {\it i.e.}, $\tilde{S}(\xi)=0$ in Eq.~(\ref{Lc-eqn_1}), the gap equation becomes $\xi$-independent and we have:
\be
\overline{L}_c=28.0981, \qquad \overline{N}_c=2.85\, .
\label{overlN}
\ee
The results of Eq.~(\ref{overlN}) are in full agreement with the recent results of [\onlinecite{Gusynin:2016som}] where the NLO corrections
have been analysed in an approximation corresponding to $\tilde{S}(\xi)=0$, {\it i.e.}, taking into account
only the NLO terms $\sim \beta$ and $\sim \beta^2$, see Appendix for details.

\section{Conclusion } 
\label{sec:conclusion}

We have studied D$\chi$SB in QED$_3$ by including $1/N^2$ corrections to the SD equation exactly and taking into account the full
$\xi$-dependence of the gap equation. Following Nash, the wave function renormalization constant has been resummed at the level of the gap equation
leading to a very weak gauge-variance of the critical fermion number $N_c$. The value obtained for the latter, Eq.~(\ref{Nc-values}), suggests
that D$\chi$SB takes place for integer values $N \leq 3$ in QED$_3$.

\acknowledgments
We are grateful to Valery Gusynin for discussions.
One of us (A.V.K.) was supported by RFBR grant 16-02-00790-a.
Financial support from Universit\'e Pierre et Marie Curie and CNRS is acknowledged.

\appendix

\begin{widetext}

\section{Resummation}
\label{app}

In this Appendix, we give some details related to the resummation procedure of Nash [\onlinecite{Nash89}] 
within our working frame. As we shall explain below, the resummation procedure consists in resumming the LO term together with part of the NLO corrections containing terms $\sim \beta^2$.
In order to do so, we consider the gap equation corresponding to Eq.~(\ref{gap-eqn-NLO-explicit_1}) in the main text and that we reproduce here for clarity:
\bea
1 = \frac{(2+\xi)\beta}{L} + \frac{1}{L^2}
\Bigl[8 \tilde{S}(\al,\xi) -2(2+\xi) \hat{\Pi} \beta + \left(\frac{2}{3}-\xi\right)\bigl(2+\xi\bigr)\, \beta^2 +
4\beta\left(\xi^2-\frac{4}{3}\xi - \frac{16}{3}\right) \Bigr]\, ,
\label{app:gap-eqn-NLO-explicit_1}
\eea
where (as displayed in the main text)
\be
\tilde{S}(\al,\xi) = \Bigl(\tilde{\Sigma}_3(\al,\xi)+2(1-\xi) \tilde{\Sigma}_2(\al) \Bigr)/8\, ,
\label{app:tS(al,xi)}
\ee
and (as found in the main text)
\begin{subequations}
\bea
&&\tilde{\Sigma}_2(\alpha=1/4) = 4\tilde{R}_1, \qquad \qquad \qquad \qquad \qquad \qquad \qquad \qquad \tilde{R}_1 = 3.7428\, ,
\label{app:tsigma2}
\\
&&\tilde{\Sigma}_3(\alpha=1/4,\xi) = \bigl(\xi^2-1\bigr)  \tilde{R}_2 -  \bigl(7+16\xi -3\xi^2 \bigr) \frac{\tilde{P}_2}{16}, ~~\qquad \tilde{R}_2 = 1.175,~~ \tilde{P}_2 = -19.28\, .
\label{app:tsigma3}
\eea
\end{subequations}
Eq.~(\ref{app:gap-eqn-NLO-explicit_1}) is the convenient starting point to perform a resummation \`a la Nash \cite{Nash89}.
In order to implement this resummation, let us first consider the integral:
\be
   \int_0^{a} \! \! \! d |k| \,
\frac{\Sigma (|k|) }{ \mbox{Max}({|k|,|p|})} {\left[\frac{\mbox{Max}({|k|,|p|})}{\mbox{Min}({|k|,|p|})}\right]}^{\lambda} ~~ ,
\label{app:SD-sigma-LO3-l}
\ee
with some arbitrary $\lambda$. Using the fact that: $\Sigma (p) = B (p^2)^{ -\alpha}$, we have:
\bea
1=\frac{1}{\Sigma (p)}
   \int_0^{a} \! \! \! d |k| \,
\frac{\Sigma (|k|) }{ \mbox{Max}({|k|,|p|})} {\left[\frac{ \mbox{Max}({|k|,|p|})}{ \mbox{Min}({|k|,|p|})}\right]}^{\lambda} 
= \left(\frac{1}{2\alpha-\lambda} + \frac{1}{1-2\alpha-\lambda} \right) =  \frac{1-2\lambda}{(2\alpha-\lambda)(1-2\alpha-\lambda)}
\, .
\label{app:SD-sigma-LO3-la}
\eea
Taking the derivative of $\lambda$ and putting $\lambda=0$, we have another important integral:
\bea
1=\frac{1}{\Sigma (p)}
   \int_0^{a} \! \! \! d |k| \,
\frac{\Sigma (|k|) }{ \mbox{Max}({|k|,|p|})} \ln \left[\frac{ \mbox{Max}({|k|,|p|})}{ \mbox{Min}({|k|,|p|})}\right] 
= \left(\frac{1}{(2\alpha)^2} + \frac{1}{(1-2\alpha)^2} \right) =  \frac{\beta}{16}\bigl(\beta -8 \bigr)\, .
\label{app:SD-sigma-LO3-lo}
\eea
So, now, we can represent the gap equation (\ref{app:gap-eqn-NLO-explicit_1}) in the form:
\be
1 = \frac{4(2+\xi)}{L\Sigma (p)}  \int_0^{a} \! \! \! d |k| \, \frac{\Sigma (|k|) }{ \mbox{Max}({|k|,|p|})}
\left\{1 + \frac{4(2-3\xi)}{3L} \ln \left[\frac{ \mbox{Max}({|k|,|p|})}{ \mbox{Min}({|k|,|p|})}\right] \right\}
+ \frac{\Delta(\al,\xi)}{L^2} \, ,
\label{app:gap-eqn-NLO-explicit_2}
\ee
where
\bea
\Delta(\al,\xi)  = 8 \tilde{S}(\al,\xi) - 4\beta\left(\xi^2 +4\xi + \frac{8}{3} + \frac{2+\xi}{2} \hat{\Pi} \right) \, . 
\label{app:Delta}
\eea
Following Nash \cite{Nash89}, the integral (\ref{app:gap-eqn-NLO-explicit_2}) may be viewed as the first two orders of the expansion of the integral (\ref{app:SD-sigma-LO3-l})
with the anomalous dimension $\lambda$ corresponding to the wave function renormalization constant:
\be
\lambda_A = \frac{\lambda^{(1)}}{L} + \frac{\lambda^{(2)}}{L^2} + ... , \qquad \lambda^{(1)}= 4 \left(\frac{2}{3}-\xi \right)\, ,
\label{app:lambda}
\ee
where $\lambda^{(1)}$ is the LO part and $\lambda^{(2)}$ the NLO one. In order to resum this contribution, we perform the following replacement:
\be
 \int_0^{a} \! \! \! d |k| \, \frac{\Sigma (|k|) }{ \mbox{Max}({|k|,|p|})}
\left\{1 + \frac{4(2-3\xi)}{3L} \ln \left[\frac{ \mbox{Max}({|k|,|p|})}{ \mbox{Min}({|k|,|p|})}\right] \right\} \to
\int_0^{a} \! \! \! d |k| \, \frac{\Sigma (|k|) }{ \mbox{Max}({|k|,|p|})}
 {\left[\frac{ \mbox{Max}({|k|,|p|})}{ \mbox{Min}({|k|,|p|})}\right]}^{\lambda_A}\, .
\ee
After this replacement, the gap equation  (\ref{app:gap-eqn-NLO-explicit_2}) takes the form:
\be
1 = \frac{4(2+\xi)}{L}  \frac{1-2\lambda_A}{(2\alpha-\lambda_A)(1-2\alpha-\lambda_A)}  + \frac{\Delta(\al,\xi)}{L^2} \, .
\label{app:gap-eqn-NLO-explicit_3}
\ee
It is convenient to multiply Eq.~(\ref{app:gap-eqn-NLO-explicit_3}) by the factor $(2\alpha-\lambda_A)(1-2\alpha-\lambda_A)$. This yields:
\be
(2\alpha-\lambda_A)(1-2\alpha-\lambda_A) = \frac{4(2+\xi)}{L} \bigl(1-2\lambda_A\bigr) + (2\alpha-\lambda_A)(1-2\alpha-\lambda_A)
\frac{\Delta(\al,\xi)}{L^2} \, .
\ee
Note that the l.h.s. can be represented as $2\alpha(1-2\alpha) - \lambda_A(1-\lambda_A)$ which leads to:
\be
2\alpha(1-2\alpha) = \lambda_A(1-\lambda_A) + \frac{4(2+\xi)}{L} \bigl(1-2\lambda_A\bigr) + (2\alpha-\lambda_A)(1-2\alpha-\lambda_A)
\frac{\Delta(\al,\xi)}{L^2} \, .
\label{app:gap-eqn-NLO-explicit_4}
\ee
From Eq.~(\ref{app:gap-eqn-NLO-explicit_4}), we see that, after resummation, $\lambda_A$, Eq.~(\ref{app:lambda}), will contribute to the gap equation up to NLO. The expression of $\lambda^{(2)}$
can be obtained from Gracey's calculations~\cite{Gracey93} and reads:
\bea
\lambda^{(2)} = - 8 \left(\frac{8}{27} + \left(\frac{2}{3}-\xi \right) \hat{\Pi} \right)\, .
\label{app:lambdaA-NLO}
\eea
Hence:
\bea
\lambda_A(1-\lambda_A) + \frac{4(2+\xi)}{L} \bigl(1-2\lambda_A\bigr) &=& \frac{4(2+\xi) + \lambda^{(1)}}{3L} +  \frac{1}{L^2} \left(
\lambda^{(2)} - {\bigl(\lambda^{(1)}\bigr)}^2 - 8(2+\xi) \lambda^{(1)}\right) \nonumber \\
&=& \frac{32}{3L} +  \frac{1}{L^2} \left(\lambda^{(2)} - 4\lambda^{(1)}
\left(\frac{14}{3}+\xi \right) \right)\, ,
\eea
which shows the complete cancellation of the $\xi$-dependence at LO. Now it is convenient to return to the standard form for the gap equation
by multiplying Eq.~(\ref{app:gap-eqn-NLO-explicit_4}) by the factor $1/[ 2\alpha(1-2\alpha)]$. This yields:
\be
1 = \frac{8\beta}{3L} +   \frac{\beta}{4L^2} \left(\lambda^{(2)} - 4\lambda^{(1)}
\left(\frac{14}{3}+\xi \right) \right) + \frac{\Delta(\al,\xi)}{L^2} \, ,
\label{app:gap-eqn-NLO-explicit_5a}
\ee
or more explicitly (as shown in the main text):
\be
1 = \frac{8\beta}{3L}  + \frac{1}{L^2}\, \Bigl [ 8 \tilde{S}(\al,\xi) - \frac{16}{3} \, \beta \, \left(\frac{40}{9} + \hat{\Pi} \right) \Bigr] \, ,
\label{app:gap-eqn-NLO-explicit_5b}
\ee
which displays a strong suppression of the $\xi$-dependence at NLO: the $\xi$-dependence exists but only through terms which are very small numerically.

We may now consider Eq.~(\ref{app:gap-eqn-NLO-explicit_5b}) at the critical point, $\al=1/4$ ($\beta=16$), which yields:
\bea
L_c^2 -\frac{128}{3} L_c - \Bigl[ 8 \tilde{S}(\xi) - \frac{256}{3}\,\left(\frac{40}{9} + \hat{\Pi} \right) \Bigr] = 0 \, ,
\label{app:Lc-eqn_1}
\eea
where $ \tilde{S}(\xi)= \tilde{S}(\al=1/4,\xi)$.
Solving Eq.~(\ref{app:Lc-eqn_1}), we have two standard solutions:
\bea
L_{c,\pm} = \frac{64}{3} \pm \sqrt{d_2(\xi)}, \qquad d_2(\xi)= {\left(\frac{64}{3}\right)}^2 + \Bigl[ 8 \tilde{S}(\xi) - \frac{256}{3}\,\left(\frac{40}{9} + \hat{\Pi} \right) \Bigr]\, . \quad
\label{app:Lcpm}
\eea
As explained in the main text, a numerical estimate of $N_c$ can be obtained by using the values of 
$\tilde{R}_1$, $\tilde{R}_2$ and $\tilde{P}_2$ in Eqs.~(\ref{app:tsigma2}) and (\ref{app:tsigma3}).
Combining these values with (as displayed in the main text): 
%
\bea
\tilde{S}(\xi=0)=\tilde{R}_1-\frac{\tilde{R}_2}{8}-\frac{7\tilde{P}_2}{128} \, , \qquad \tilde{S}(\xi=1)= -\frac{5\tilde{P}_2}{32} \, , \qquad
\tilde{S}(\xi=2/3)=\frac{\tilde{R}_1}{3}-\frac{5\tilde{R}_2}{72} - \frac{49 \tilde{P}_2}{384}\, ,
\nonum
\eea
together with the value of $\hat{\Pi}=92/9 - \pi^2$, yields, for $L_c(\xi)$ and $N_c(\xi)$:
\begin{subequations}
\bea
&&L_c^+(0)=30.44, ~~ L_c^-(0)=12.22, \quad L_c^+(2/3)=29.98, ~~ L_c^-(2/3)=12.69 \quad L_c^+(1)=29.69, ~~ L_c^-(1)=12.98\, , \qquad 
\label{app:Lc-res} \\
&&N_c^+(0)=3.08, ~~ N_c^-(0)=1.24, \quad N_c^+(2/3)=3.04, ~~N_c^-(2/3)= 1.29 \quad N_c^+(1)=3.01, ~~N_c^-(1)=1.31 \, , \qquad
\label{app:Nc-values}
\eea
\end{subequations}
where the ``$-$'' solutions are considered as unphysical. The ``$+$'' solutions are the ones presented in the main text.
Notice that if we neglect the rest in Eq.~(\ref{app:Lc-eqn_1}), {\it i.e.}, we assume that $\tilde{S}(\xi)=0$ in Eq.~(\ref{app:Lc-eqn_1}), the gap equation becomes $\xi$-independent 
and we have (as given in the main text):
\be
\overline{L}_c=28.0981\, , \qquad \overline{N}_c=2.85 \, .
\label{app:Nc-values-tS=0}
\ee
The result of Eq.~(\ref{app:Nc-values-tS=0}) is in full agreement with the recent results of Ref.~\cite{Gusynin:2016som} 
 where the NLO corrections have been analysed in an approximation corresponding to $\tilde{S}(\xi)=0$, {\it i.e.}, taking into account
only the NLO terms $\sim \beta$ and $\sim \beta^2$ in the rhs of Eq.~(\ref{app:gap-eqn-NLO-explicit_1}).

\end{widetext}

\end{fmffile}

\end{document}